\documentstyle[prl,aps]{revtex}

\begin{document}
\draft
\title{Enhanced stability of the square lattice of a classical bilayer Wigner 
crystal}

\author{I. V. Schweigert \cite {d:gnu},
V. A. Schweigert \cite {*:gnu} and   F. M.  Peeters 
\cite {f:gnu}}
\address{\it  Departement Natuurkunde, Universiteit 
Antwerpen (UIA),\\
Universiteitsplein 1, B-2610 Antwerpen, Belgium}
\date{\today}
\maketitle

\begin{abstract}  
The stability and melting transition of  a single layer and a 
bilayer crystal consisting of  charged particles  interacting 
through a Coulomb  or a screened Coulomb potential is 
studied using the Monte-Carlo technique. 
A new  melting criterion is formulated which we show to 
be  universal  for bilayer  as well as for single layer 
crystals   in the case of (screened) Coulomb, 
Lennard--Jones and $1/r^{12}$ repulsive inter-particle  
interactions. 
The melting temperature for the five different lattice 
structures of the bilayer Wigner crystal is obtained,
and a phase diagram is constructed 
as a function of the interlayer distance. We found the
surprising result that the 
square lattice has a substantial larger melting temperature 
as compared to the other lattice structures. 
This is a consequence of the specific topology of the 
defects which are created with increasing temperature and which have a larger
energy as compared to the defects in e.g. a hexagonal lattice.
\end{abstract} 

\section {Introduction}
Phase transitions and in particular the melting transition
is one of the most fundamental problems
 in condensed matter and statistical physics.
The most intensively studied model system
 is the one consisting of charged particles
 with a pure spherical symmetric
 repulsive inter--particle interaction 
(see e.g. \cite {Kalman}).
 Examples are the electron Wigner solid, colloidal spheres and dusty plasmas.
 After the discovery of the Wigner crystallisation
 of electrons above the surface of liquid helium by
 Grimes and Adams \cite {Grimes} the interest
 in the melting of low dimensional systems has increased. 
Recently the layered close-packed crystalline structure 
  and their structural transition  were observed in experiments
with dust rf discharge 
\cite {Hayashi} and with ion layered crystals in ion traps \cite {Dubin98}.
Motivated by the theoretical works
of Nelson, Halperin 
\cite{Nelson}, and Young \cite {Young} who  developed a theory for a two stage 
continuous melting of a  two dimensional (2D) crystal and which was 
based on 
ideas of Berenzinskii \cite {Berenzinskii}, Kosterlitz and 
Thouless \cite{Kosterlitz},
several experimental 
\cite{Murray,Murray1,Seshadri,Kusner} and theoretical 
studies 
\cite{Morf,Saito,Strandburg,Beda,Zollweg,Lee,Naidoo,Weber,Li,Sadr,Kleinert,Chui,peetersnew} 
were devoted to  the  melting  transition of
mainly single layer crystals. 
In this case the  
hexagonal lattice is  the most energetically 
favored structure for potentials of the form $1/r^n$ \cite {Gann,Broughton}. 
The effect of the dimensionality of the system on the
 melting transition is another important issue
 which has been investigated recently. An important  subsystem in this
 field is the bilayer system consisting of two parallel
 layers of charged particles with a pure repulsive inter-particle interaction.
The interlayer distance is an additional variable with which the
effective interparticle interaction can be altered.

While the classical single layer
 can crystallise only in a
 hexagonal lattice it was found
 that the bilayer system exhibits a much
 richer structural phase diagram.  
Previously, the different types 
of lattices and structural transitions in a multilayer crystal 
at zero temperature with parabolic lateral
 confinement was analysed in \cite  {Dubin,Schweigert,Totsuji}.
Different classes of lattices of the double-layer crystal
 were specified in \cite {Falko}
 as function of the interlayer spacing and electron density.
 A systematic study of the stability and the phonon
 spectra of these  crystal phases 
 was presented in Ref.\cite {Goldoni}.
The collective modes of the corresponding bilayer liquid state was
recently studied in \cite{kalman_new}. The extension of the classical bilayer
model to the finite vertical coupled classical dot system was
presented in Ref.\cite{partoens} where a rich collection of structural first
and second order phase transitions was found.
 The quantum bilayer system was investigated by several groups 
\cite {Swierkovski,Goldoni95,Rapisodo,Voltchinov}.

Here we will 
concentrate on the melting of the classical bilayer crystal and study
 the influence  of the different crystal structures on
 the melting temperature. The  equilibrium states of 
a bilayer crystal at different temperatures ($T$) will be investigated
 using   the 
Monte Carlo (MC) simulation technique. Preliminary resuls of this
study were published in \cite{schweigert_new}.

The melting of the classical bilayer crystal was
 studied in Ref. \cite {Goldoni} using the
 Lindemann criterion where the mean square
 displacement was calculated within the harmonic approximation.
It is well-known that the harmonic approximation can only
 give a rough estimate of the melting temperature.
 For example, for the single electron layer using the
Kosterlitz--Thouless--Halperin--Nelson--Young (KTHNY) theory and
 inserting the Lam\'e coefficients of the $T=0$ lattice
 one obtains $\Gamma =78$, while inclusion of non linear effects,
 using Monte-Carlo simulations \cite {Morf},
 results into $\Gamma =125$ which compares to
 the experimental result $\Gamma =137\pm 15$ 
(see e.g.\cite {Peeters}). This is the motivation to go
 beyond the harmonic  approximation and to investigate the
 importance of non linear effects on the melting temperature. 
Using the Monte Carlo simulation technique
has the added advantage that information can be obtained about the 
topology of the defects which are responsible for melting.

The present paper  is organised as follows. The model system and the
numerical approach is given in Sec.II. In Sec. III  we 
construct the solid--liquid phase diagram for Coulomb 
bilayer crystals. In Sec. IV the influence of screening on 
melting is studied and
we formulate a new criterion
 for the melting temperature. In Sec. V we show that it gives
accurate results for the melting temperature when we use the new
melting criterion and the harmonic approximation.
 The existence of hysteresis during melting
  of the 2D single layer crystal is considered in Sec. VI. 
The topology of defects in  the 
hexagonal type crystal is considered in Sec. VII. The 
gallery of defects for the square  bilayer crystal is 
presented in Sec. VIII. Our results are summarised in Sec. 
IX.

\section {The model system and numerical approach}
In the present study we limit ourselves to infinitely thin bilayers with equal 
density, $n/2$, of charged particles in the two layers.
In the crystal phase the particles are  arranged in two parallel layers  in 
the $(x,y)$--plane which
are a distance $d$ apart in the 
$z$--direction. The lattice structure in the top and bottom layers are the 
same, but the particle sites in  opposite layers are staggered 
(i.e. closed-packing system).
A single layer crystal is a limiting 
case of a bilayer crystal with $d=0$ and particle 
density $n$.
  
We  assume that the particles interact through an isotropic  
Coulomb  or screened repulsive potential
\begin{equation}
\label{eq1} 
V({\vec {r}}_i, {\vec {r}}_j) =\frac {q^2}{\epsilon 
\mid {\vec {r}}_i-{\vec {r}}_j\mid}
\exp (-\kappa {\mid {\vec {r}}_i -{\vec {r}}_j\mid}),
\end{equation} 
where $q$ is the particle charge,  $\epsilon $ is   the  
dielectric constant of the medium  the particles are moving 
in,
  $\vec {r}=(x,y,z)$
 the position of the particle with $r=\mid \vec {r}\mid $,
and $1/\kappa $ is the screening length. 
 The type of lattice symmetry  at $T=0$ depends on the 
dimensionless parameter $\nu =d/a_0$, where   
$a_0 =1/\sqrt {\pi n}$  is the mean 
interparticle distance. 
For the classical Coulomb system 
($\kappa =0$) at $T\neq 0$ there are two 
dimensionless parameters $\nu $  and $\Gamma =q^2/\epsilon a_0 
k_BT$ which determine  the state of the system. The 
classical Yukawa system ($\kappa >0$) at $T\neq 0$ is 
characterised by three independent dimensionless 
parameters: $\nu $, $\Gamma $ and $\lambda =\kappa 
a_0$. Below we measure the temperature in units of 
$T_0=q^2/\epsilon a_0k_B$ and the energy in $E_0=k_BT_0$. 
In addition, we will also consider the melting of a single 
layer crystal with a Lennard-Jones ($1/r^{12}-1/r^6$) 
inter-particle potential  and with a $1/r^{12}$ repulsive potential as 
well.

In our simulations the density of particles $n$ remains the same for
 all interlayer distances $\nu $ while the interlayer distance changes. 
The initial symmetry of the 
structure is set by the primitive vectors, the values of 
which  are derived from a calculation of the  minimal 
energy configuration for fixed $\nu $ at $T=0$ using
 the Ewald method \cite {Ewald}. 
In our numerical calculations  we took  a fragment
 of  lattice (ranging from $N=228$ to $N=780$ particles) and used  periodic 
boundary conditions. 
 
The structure and potential energy of the system
at $T\neq 0$ are found by the standard  Metropolis algorithm 
\cite {Metropolis} in which at
  some temperature the next simulation state of the system
 is obtained by a random displacement of one of the  particles.
 If the new configuration has a smaller energy
 it is accepted and if the new energy is
 larger the configuration is accepted if 
 $\delta >r$, where the probability
$\delta =\exp (-\Delta E/k_BT)$ with  $\Delta E$ the increment in the energy and
$r$ is a random number between $0$ and $1$. 
We allow the system to approach its  
equilibrium state at  some temperature $T$, after 
executing $(10^4\div 5\times 10^5)$ `MC steps'. 
The potential energy  at $T\neq 0$ is found by summation over all 
particles and their periodic images using the Ewald method \cite {Ewald}. 

\section {Solid--liquid phase diagram}

In \cite {Goldoni} it was 
found  that the bilayer Coulomb 
crystal at $T=0$ exhibits  five different lattices  as  
function of the interlayer distance: 
 $\nu <0.006$--hexagonal, $0.006<\nu 
<0.262$--rectangular, $0.262<\nu <0.621$--square,
$0.621<\nu <0.732$--rhombic,  and $\nu 
>0.732$--hexagonal.

It should be noted that
the behaviour of the system near the melting point is essentially non-linear
 and therefore the melting temperature obtained in Ref. \cite {Goldoni} 
has only qualitative predictive power. Therefore,  we revise the melting phase 
diagram and present a more accurate calculation of it, by performing  MC 
simulation of the melting transition. 

We use different criteria to find the 
critical melting temperature. 
The parameter $L_*=<u_*^2>/a_0^2$, where $<u_*^2>$  is the mean square 
displacement of the particles from their equilibrium positions 
was introduced in \cite {Lindemann}  
to characterise the system order. It is well--known that $L_*$ diverges
 for a 2D system. We will use a modified 
parameter $L$ as introduced in Ref.\cite{Beda}, 
which enables us to minimize the system size effect which 
gives significant contributions in the usual parameter 
$L_*$. For the modified parameter $L=<u^2>/a_0^2$, $<u^2>$ is 
defined by the difference in the mean square displacement 
of neighboring particles from  their initial sites 
${\vec {r}}_0$ and can be written as
\begin {equation}
\label{eq2}
<{u}^2> = \langle 
\frac{1}{N}          {\sum\limits_{i=1}^{N}}
\frac{1}{N_{nb}}{\sum\limits_{j=1}^{N_{nb}}}
(({\vec {r}}_i - {\vec {r}}_{i0})-
({\vec {r}}_j-{\vec {r}}_{j0}))^2\rangle ,
\end {equation}
where $<>$ means averaging over MC steps, the index $j$ denotes the $N_{nb}$ 
nearest neighbours of  
particle $i$ in the upper or in the bottom layers. For  the 
hexagonal lattice $N_{nb}=6$, while for 
rectangular, square and rhombic lattices 
$N_{nb}=4$. 

For sufficiently low temperatures the particles
exibit  harmonic oscillations around their $T=0$ equilibrium position, 
and the oscillation amplitude increases 
linearly with temperature, which leads to a linear increase 
of $L$ with $T$. Figs.~1--3(a) show 
the modifies parameter $L$  as a function 
of the reduced temperature $T/T_0$: i) for the
hexagonal single layer 
crystal ($\nu =0$) (see Fig.~1(a)), ii) for the bilayer crystal with a square  
lattice ($\nu =0.4$) (see Fig.~2(a)), and iii) for a hexagonal type
 bilayer crystal  
($\nu =0.8$) (see Fig.~3(a)).
At higher 
temperatures non-linear effects are important and
 $L$ becomes a nonlinear function of  
$T$. The particle oscillation amplitude  increases faster 
 than linear with $T$, but the system is not melted,
 since the particles display still an ordered structure. The rapid 
increase of $L$ with $T$ is a manifestation of the 
anharmonicity of the motion of the particles.
Melting occurs when $L$ increases very sharply with $T$. 
In our previous study \cite{Goldoni} we assumed that melting occurs when
$L \sim 0.1$ as was 
obtained from numerical simulations of melting in a single layer crystal 
\cite {Beda}. 
From the present study we find that $L \sim 0.1$ is not a good criterion
for melting, except for the case $\nu =0$ while for $\nu =0.4$
 and $\nu =0.8$ the system is still ordered when $L\sim 0.1$. Note, that we 
defined 
$L=<u^2>/a_0^2$, where $a_0$ is the mean inter-particle  distance for a single 
layer crystal with density $n$, for all interlayer distances $\nu $.
In the case when $\nu \rightarrow \infty $ and the inter--layer particle 
interaction is negligible small, we obtain $L\sim 0.2$ at melting because now 
the particle density in each layer is $\sqrt {2}$ times smaller.
 
As a second independent parameter from which we determine the melting
temperature we calculated the height of the first peak in the pair 
correlation function as function of temperature. The pair correlation function 
is calculated for the particles in one of the layers (see Fig.~4). This is  
shown in Fig.~4 for an interlayer separation $\nu =0.4$ and three different 
temperatures. The height of the first peak in $g(r)$ is plotted as function of 
temperature 
in  Figs.~1-3(b) for $\nu =0,  0.4,  0.8$, 
respectively. The value of the first peak of the pair 
correlation function decreases with increasing temperature, 
and exhibits a quick decrease at $T=T_{mel}$. This is most pronounced for the 
case of a 
single layer crystal (Fig.~1(b)). 
Even in the liquid phase the pair correlation function still has a peak
structure due to the local inter-particle correlation.  In Fig~4 the state  of 
the system just before melting is denoted by the doted  curve and the dashed 
curve refers 
to the state after melting. 

In the crystalline state the  potential energy of the system  
increases almost linearly with temperature and only a weak non linear
$T$-dependence is observed. 
At larger temperature 
a quick rise in the energy is clearly seen in 
Figs.~1-3(c) denoting the beginning of melting   and is related to 
the  unbinding of dislocation pairs, which we will discuss 
below.  At melting the potential energy has a very steep rise.
The heat capacity of a 2D structure is determined by
 $C=(d\varepsilon /dT)_{V=const}$,
 where the total energy $\varepsilon =k_BT_0+E$ includes the kinetic and 
potential energy. The heat capacity of the crystal  and the liquid 
phases near melting are very close to each other in two dimensions, and the 
first derivative of the energy with 
respect to T taken in the crystal phase and the liquid phase 
are practically the same. The potential energy
 decreases with increasing $\nu $ which is a consequence of the
decreasing  inter-layer Coulomb interaction. 
Note, that the potential energy difference between  the liquid and solid phases 
near melting for the square bilayer crystal at  
$\nu =0.4$ is of size 
$\delta _e=0.71 \times 10 ^{-2}k_BT_0$ which is about a factor 2 
larger than for a hexagonal lattice, i.e.  at $\nu =0$,  
$\delta_e=0.39\times 10 ^{-2}k_BT_0$,   and at $\nu =0.8$,  
$\delta _e=0.31\times 10 ^{-2} k_BT_0$.   Moreover, as follows
 from the temperature behaviour of the parameter $L$, 
the peak of the pair correlation function and the potential energy, the 
square lattice turns out to exibit 
a substantial higher melting temperature,
 and consequently is more stable against thermal 
fluctuations, than the hexagonal one.  The  enhanced stability of 
the square lattice can be understood from an analysis of the defects which are 
formed during melting and which  will 
be discussed in Secs. VII and VIII. 

In order to determine the type of melting transition and the 
melting point the bond--orientational  and  translational 
correlation functions are  calculated 
\cite {Beda,Naidoo,Li} which  sometimes can also be 
obtained 
experimentally \cite {Murray,Kusner,Pieper}.
But our finite fragment of 780 particles (with 
periodic boundary conditions) is too small for a reliable 
analysis of the asymptotic decay of the density correlation 
functions. Therefore,   
we calculate the  bond--angular order factor $G_\theta $
which originally was
introduced by Halperin and Nelson \cite {Halperin}. For each layer ($k$) 
we define the bond-angular order factor
\begin {equation}
\label{eq3}               
G_\theta ^k = \langle 
\frac{2}{N}
            {\sum\limits_{j=1}^{N/2}}
\frac{1}{N_{nb}}{\sum\limits_{n=1}^{N_{nb}}}
\exp (iN_{nb}\theta _{j,n}) \rangle,
\end {equation}
and the translational order factor
\begin {equation}
\label{eq4}
G_{tr}^k= \langle
\frac{2}{N}
{\sum\limits_{j=1}^{N/2}}
\exp (i\vec G\cdot ({\vec {r}}_i-{\vec {r}}_j)) \rangle ,
\end {equation}
where $\theta _{j,n}$ is the angle between some 
fixed axis and the vector which connects the $j$th particle 
and its nearest $n$th neighbour, and $\vec G$ is a reciprocal--lattice vector. 
The latter we took equal to the smallest reciprocal lattice vector.
The total bond-angular order factor of the bilayer crystal is defined as 
 $G_\theta =(G_\theta ^1+G_\theta ^2)/2$ and similar for  $G_{tr}$. 

These order factors are shown 
in Figs.~1-3(d) as function of temperature.
For low temperatures the factors $G_{\theta}$ (solid circles) and $G_{tr}$ (open 
circles) decrease linearly 
with $T$, and when  $G_\theta $ reaches $0.45$, both 
order factors drop quickly to zero at the same temperature 
as seen in Figs.~1-3(d).
From the behaviour of the order factors we can  
derive the temperature at which order is lost in the system. 
Our numerical results show that the bond-order angular order factor: 
1) decreases linearly with increasing temperature, except very close
 to the melting temperature, where it decreases faster, and 
2) it drops to zero just after it 
reaches the value of 0.45. 

In the phase diagram of Fig.~5 we show the melting 
temperature as a function of  $\nu $, which is 
the interlayer distance for fixed particle density. All 
criteria mentioned above gave the same temperature   $T_{mel}$.
For $\nu =0$ we obtained the  
critical $\Gamma^*  =132$, resulting in $T/T_0=0.0076$. 
This critical value was first measured in Ref.\cite {Grimes} and found to be 
$137\pm15$. Later experiments \cite {gamma_e} 
and simulations \cite {gamma_t}
gave critical $\Gamma ^*$ within this range. 
As seen in Fig. 5 the hexagonal (I and V),  the rectangular (II) and rhombic 
(IV)
lattices melt at  almost the same temperature. This 
implies that the inter-layer correlation does not strongly 
influence the melting temperature, although it determines 
which lattice structure is stable and has the lowest energy. 
Note, however that the melting temperature for phase IV and V are slightly
smaller than the  corresponding temperature for phases I and II in
the given $\nu$ range. We checked that in the limit $\nu\rightarrow 0$
$T_{mel}/T_0\approx 0.0076/\sqrt{2}\approx 0.0537$, which is reached
for $\nu=2\div 3$.

The general shape of the phase diagramm agrees with the one obtained earlier
\cite{Goldoni} using the Lindemann criterion where the mean square
displacement was calculated within the harmonic approximation. But
there are several important differences: 1) the absolute value of the
melting temperature was underestimated in Ref.\cite{Goldoni} by about a factor of 
$2$; 2) at
the second-order structural phase transition points the melting
temperature becomes zero wich was a consequence of the softening of a
phonon mode at $T=0$. With the present Monte-Carlo simulations we
find $T_{mel}\neq 0$ which is due to the importance of non-linear
effects; 3) in Ref.\cite {Goldoni} the maximum melting temperature in phase II 
was larger
than the one in phase III (square lattice) which is opposite to what
is found in the present calculation; and 4) the melting temperature
in phase IV was smaller than in phase V for $\nu<1$, which is
opposite to the present results.

For the square bilayer crystal (III)  the melting temperature
increases with rising $\nu $ and 
only for $\nu >0.4$ we found that 
$T_{mel}$ starts to decrease with increasing $\nu $. It is surprising that the 
square lattice bilayer crystal has the 
maximal melting temperature $T_{mel}=0.01078T_0$ which exceeds the critical 
temperature of the single layer crystal $T_{mel}=0.0076T_0$, where the particle 
density is two times larger. 
The most stable square lattice bilayer occurs at $\nu =0.4$ which results into 
the critical coupling parameter $\Gamma ^*=187$. In the present paper we will 
not discuss the temperature 
induced structural phase transition, but limit ourselves to 
the melting transition. We denote in Fig.~5 the phase 
boundaries between the different crystal phases by vertical dotted lines. 
The analysis of the stability of the 
classical bilayer crystal was carried out in 
Ref.\cite{Goldoni} using the harmonic approximation
 and it was found that near the structural phase transition the melting 
temperature
was strongly reduced due to
the softening of a phonon 
mode which  may lead to 
a re--entrant behaviour.
  No such behaviour was found in the present Monte-Carlo simulaion.
 Near the structural phase transition we found
 deformed lattice structures with a long wavelength deformation. At present 
we cannot draw  any definite conclusions from it and we 
 need to simulate larger crystal fragments 
during more Monte-Carlo steps. But this observation does 
not influence our conclusions concerning the melting behaviour.

\section {Melting of bilayer crystal
 with screened inter--particle interaction and formulation of a new
melting criterion.}

Symmetry of the lattice of the bilayer crystal at $T=0$
 might depend not only on the interlayer distance, but also
 on the type of the inter-particle interaction.
 The phase diagram corresponding to the lowest  energy lattice 
structures for a bilayer crystal at $T=0$  with screened Coulomb 
inter-particle interaction is shown in Fig.~6 as function of 
the screening length $\lambda $.
Notice that the phase boundaries depend only weakly on 
the screening length. On the other hand,  the melting 
temperature   reduces considerably
 with screening as is apparent from Fig.~5
in the case $\lambda  =1$ (solid circles) and $\lambda =3$ (open triangles).
The melting temperature of  a single 
layer crystal with Coulomb (i.e. $\lambda =0$) interaction is 
$T_{mel}=0.0076T_0$, with screened one  is 
$T_{mel}=0.0066T_0$ for $\lambda =1$, and for $\lambda =3$  it is 
$T_{mel}=0.0035T_0$. Note, that
 the bilayer crystal with screened inter--particle interaction 
has the same qualitative melting curve as the 
Coulomb bilayer crystal (see Fig.~5). 

For the case of a screened Coulomb interparticle interaction
 ($\lambda =1$) the change of the potential
 energy with temperature as referred
 to the $T=0$ result is shown in Fig.~7(a) for two different inter-layer 
distances. Notice that the
potential energy initially increases linearly with temperature but then near the 
melting temperature it rises quickly 
within a very narrow temperatures range. It is apparent that
 the square 
lattice bilayer crystal ($\nu = 0.4$) has an enhanced melting temperature
 as compared to the hexagonal lattice (i.e. $\nu =0$).
The system looses  order  just after the 
bond--angular order factor reaches the value $0.45$ as is 
apparent from Fig.~7(b).
Also, for the screened Coulomb interaction
 the square  bilayer lattice 
 exibites an enhanced melting temperature.

From the present numerical results for $G_{\theta}$ for a screened Coulomb
interaction and from previous section
 for a pure Coulomb interaction for single
and bilayer lattices we formulate 
 a new criterion for melting which we believe
is universal: {\it melting occurs when the bond-angle correlation factor 
becomes $G_{\theta} \sim 0.45$}. 
We also checked the validity of this criterion 
for a single layer crystal with a Lennard-Jones $V= 
1/r^{12}-1/r^6$   and a repulsive $V = 1/r^{12}$ 
interaction potential.  The results for the translational and the
 bond--angular factors  for single and bilayer
 systems and for different inter-particle interactions
 are shown in Fig.~8 as function of the ratio $T/T_{mel}$.
These results  
 confirm the universality of the proposed criterion. It is valid for
single and bilayer systems and we believe it is also  independent of the 
functional form of 
the
inter-particle interaction. 
 
\section {Harmonic approximation}

Given the fact that the bond-order angular order factor decreases  almost 
linearly with 
increasing temperature, and 
melting occurs when the bond--angular 
order factor equals $0.45$ we will calculate the melting 
temperature within the harmonic approximation.
Therefore, we reformulate the
 definition of the angular-bond order factor
 within the harmonic approximation in terms of 
 eigenfrequencies and eigenvectors of the $T=0$ phonon spectrum.
Previously such a method was used 
in Ref.\cite {Schweig95} in the calculation of the modified
Lindemann parameter for a finite cluster and in Ref.\cite{Goldoni}
for the bilayer system.
   In the present paper we
apply the same  algorithm in the calculation
 of the bond angular order parameter.
We consider a finite crystal fragment with  periodic boundary conditions
and diagonalize numerically the corresponding Hessian matrix. The
theoretical background can be shortly described as follows. Any
thermodynamically averaged observable which is a
functional of the particles coordinates $\vec r_i$ can be written
as
$$G=\int G(\vec R)exp(-U(\vec R)/T)d\vec R,$$
where $\vec R=(\vec r_1, \vec r_2, ...,\vec r_N)$ is a multidimensional
vector describing the system coordinates and $U$ is the potential energy.
Within a standard linear approach we expand  both $G(\vec R)$ and
$U(\vec R)$ near the equilibrium state $\vec R=\vec R_0+\vec \xi$, where
$\partial U/\partial \vec R |_{\vec R=\vec R_0}=0$, and obtain
$$G=\int \frac{1}{2}\frac{\partial ^2G}{\partial \xi_i\partial \xi_j}
\xi_i\xi_j
exp\left (-\frac{1}{2T}
\frac{\partial ^2G}{\partial \xi_i\partial
\xi_j}\xi_i\xi_j\right)d\vec \xi.$$
The multidimensional integration can be performed after
diagonalisation of the Hessian matrix which has the elements
$$A_{ij}=\frac{\partial ^2G}{\partial \xi_i\partial\xi_j},$$
and after introduction of the normal mode coordinates.

Thus within the harmonic approach the angular--bond order factor
is given by the following expression
\begin{equation}
\label{eq5}
G_{\theta }=1-\frac{T}{2N^2}
\sum_{n=1}^{N}\sum_{k=1}^{N}\sum_{i=4}^{2N}
\left(({\vec r}_n-{\vec r}_{n1})\times 
({\vec A}_{n,i}-{\vec A}_{n1,i})
-({\vec r}_k-{\vec r}_{k1})\times 
({\vec A}_{k,i}-{\vec A}_{k1,i})\right)^2 / \omega_i^2 ,
\end{equation}
where $\omega _i$, $\vec A_i$ are the eigenfrequencies 
and the
eigenvectors of the Hessian matrix formed by the second 
derivative of the
potential energy; the indices $n1$, $k1$ refer to the 
nearest neighbours
of the particles $n$, $k$, respectively, where the summation is 
performed over all particles.
The first three eigenmodes have $\omega_i=0$ and correspond to
the center-of-mass motion and rotation of the system as a whole
which we removed from
the summation in Eq. (5).
The melting temperature is then obtained by 
taking $G_{\theta }$ equal to the value $0.45$.
The obtained results are shown in Fig.~9 by the solid curves for the 
square lattice (Fig.~9(a))
 as function of the interlayer
 distance and the hexagonal (Fig.~9(b)) lattice ($\nu =0$) as function of the
screening 
length. Thus  using our new criterion together with the harmonic
approximation reproduces 
$T_{mel}$ rather well and agrees with our MC 
calculations within $10\%$ for $0.35<\nu<0.55$ in the case of the
square lattice. Near the structural transition the harmonic
approximation underestimates the melting temperature due to the
softening of the $T=0$ phonon modes (see Ref.\cite{Goldoni}).

\section {Is there hysteresis at melting?} 

One of the most interesting questions in the field of phase transitions is 
whether or not there is a hysteresis behaviour at 
melting and freezing  of the crystal structure.  Even if such 
an hysteresis is found it can be an artefact of
the numerical simulation arising from the fact 
that the equilibrium state of the system is not reached.
To shed light on this problem we studied more carefully the melting of the 
single layer crystal in the temperature range where a rapid 
increase of the potential energy was found $T_1<T<T_3$, (see Fig.~1(c)).  
We took very small temperature increments of  
$0.001T_{mel}$. 
At the temperature $T_1=0.007560T_0$, which is just below the  
melting point, during numerical annealing  of the system 
which lasts for $5 \times 10^5N$ MC steps, the potential energy 
practically remains the same and changes only with 
$3\%$ of  the total energy jump $\delta _e$. Note, that
 the Monte-Carlo `time' step is about a 
factor 10 larger than that in typical molecular dynamic 
calculations. At the melting temperature 
$T_3=0.007565T_0$ the energy rises rapidly during  $10^4$ 
MC steps, and then the system remains in the liquid 
state during $5\times 10^5$  MC steps. We 
also found such a state of the system at the intermediate 
temperature $T_2=0.007562T_0$ when the system switches 
from solid to liquid and back as function of the simulation time. 
The system exhibits a network of grain boundary 
defects which makes it possible for the system to get back to a defect-free 
structure at the same temperature
(see Fig.~10). Such a type of equilibrium behavior of the
 system at the critical point was found earlier by Morf \cite {Morf}.
The 
potential energy of the system at this temperature and the 
order factors $G_{\theta }$, $G_{tr}$ are shown in Fig.~11 
as function of the `MC time'. As function of time the system  switches between 
the solid and liquid state where it looses and regains translational and 
orientational order. 
The existence of this state 
shows that there is a temperature range 
(at least  for a crystalline fragment of 448 particles) in which 
the crystalline  and liquid states have equal 
probability to 
occur and, consequently, no hysteresis  occurs in the 
Coulomb single layer crystal. It should be noted
 that the potential energy changes very rapidly, in an extreme narrow range of 
temperatures. Whether or not this rapid rise is continuous or discontinuous is 
not yet clear.

\section {Defects in the hexagonal lattice.}

In this section we consider the topology of the defects which are created 
during melting of a single layer crystal. 
At low temperature the probability of 
defect creation is small and the lattice remains perfect.  
With rising $T$ the energy increases and the system can overpass some potential 
barrier and
we found that during the MC simulation 
the system transits from one 
metastable state to another.  The different metastable states differ by the 
appearance 
of isomers 
in the crystal structure which appear with 
different probabilities. In order to find these isomers
  in the equilibrium state of our system we choose, during
 our MC simulation,  a number of instant particle configurations.
 Then we quickly freeze each instant particle configuration
which enables us   to exclude the 
`visual' oscillation effect, i.e. the displacements of the particles
from their lattice configurations due to thermal fluctuations. 
After 
freezing each configuration we find the topology and energy of the 
isomer and the bond-angular and the translational order 
factors of this crystal structure. 

The isomers at $T_1=0.00756T_0$, which is a temperature 
just before melting,  at $T_3=0.00765T_0$ which is  just above
 melting, and at the intermediate temperature $T_2=0.007562$  (see Fig.~1(c)), 
were obtained. 
Each point in Fig.~12(a) represents one configuration containing 
an isomer in a single layer ($\nu =0$). 
Similar  results are shown in Fig.~12(b)
 for the square lattice bilayer system ($\nu =0.4$).

With rising temperature
first, point defects such as a
 `centered vacancy' and a
 `centered interstitial' are created at $T\leq T_1=0.00756T_0$. 
These calculated defects as obtained by freezing 
 the instant particle configuration 
to zero temperature are shown in Fig. 
13(a,b).  Point defects  appear in pairs in our MC 
calculations, which are a consequence of the periodic 
boundary condition.
The total energy of a non bounded pair consisting of a  
`centered interstitial'  and a `centered  vacancy' is  
$E=0.29k_BT_0$. 
We checked the accuracy of this value on 
fragments of 512 and 780 particles which contain only one 
pair of  `centered interstitial' and  `centered vacancy'. In both 
cases  the minimum energy  of a pair of uncorrelated  point 
defects  was the same within $1\%$.  The quartet of bound 
disclinations (two sevenfold and two fivefold) also appear
at this  temperature (see Fig.~5(a) of Ref.\cite{schweigert_new}) and  has an energy 
$E=0.285 k_BT_0$.  
The point defects and  
disclinations bound into a quartet
have only negligible effect on 
$G_\theta $ and $G_{tr}$ (see Fig.~12(a), group 1).
The order factors of the `cold' configurations (freezed to 
$T=0$)   containing a `centered interstitial' and a `centered 
vacancy' equal  $G_\theta =0.8\div 0.9$ and 
$G_{tr}=0.85\div 0.95$, (see Fig.~12(a)). 
It should be noted that in spite of prolonged annealing of 
the system during $5\times 10^5$ MC steps at a 
temperature $T_1=0.007560T_0$, which is just below melting,
 we did not find more complex isomers than point defects and 
quartets of disclinations.

At the temperature $T_3=0.007565T_0$, which is  
 the melting temperature, (see Fig.~1(c)) we find
pairs of disclinations which are created by the 
dissociation of disclinations which were bound into a quartet (see Fig.~5(b)
of Ref.\cite{schweigert_new}).
 Pairs of disclinations can be found in an 
energy range of $(0.216\div 0.285)k_BT_0$ (group 2 in Fig.~12(a)).
 These defects 
destroy the translational order of the system, but conserve
the bond--angular order. The energy range of $(0.29\div 
0.62)k_BT_0$ (group 2) refers to multiple disclination pairs.  Within this 
range of energy  
the bond--angular order factor remains larger than the 
translational one. For the energy range $(0.62\div 1)
k_BT_0$ (group 3)
there are  pairs of disclinations aggregating into 
chains, which are nothing else then grain boundaries (see 
Fig.~10) which have the effect of rotating one part of the 
crystal with respect to another. In this case, both order 
factors become very small.
These defects occur in the system at intermediate 
temperature $T=T_2$ when the system exhibits the remarkable 
property of switching between a 
quasi--liquid state and a crystalline state  during our 
numerical simulation. These defects change the 
energy  substantially, but allows the system to come back to a quasi-ordered 
state (see Fig.~11).
At temperature $T>0.008T_0$ both order factors 
become small and the system transits to an isotropic fluid. 

\section {Topology of defects in the square bilayer 
crystal.} 

In order to understand why the square lattice bilayer crystal  has a 
considerable larger melting temperature as compared to e.g. the hexagonal
lattice we investigated 
the various isomers which are created in the temperature interval $T_1<T<T_2$
(see Fig.~2(c) for the location of the temperatures). 
For bilayer crystals the crystal structure and the topology 
of the defects is viewed as being composed by the top and 
the bottom staggered layers. Here, we took as an example $\nu 
=0.4$ which has the largest  melting temperature, i.e.  
$T_{mel}=0.01078T_0$. Note, that  the energy of the 
defects which occurs in the square lattice depends on the 
interlayer distance. 
The scenario of defect formation is the same as in the case of a single layer 
crystal. First, for $T\leq T_1=0.01076$
point defects are formed. These are the `twisted bond' which is shown in 
Fig.~14(a) and has the energy $E=0.23
k_BT_0$, a `twisted triangle' and a `twisted square' which are 
shown in Fig.~14(b,c), and have an energy $E=0.24
k_BT_0$ 
and $E=0.26 k_BT_0$, respectively (group 1 in Fig. 12(b)). Each of these defects 
can 
appear separately.  

The next energy group includes  the 
`vacancy' and the `interstitial' point defects which appear  in pairs, see 
Fig.~14(d,f) (group 1).  The energy of a pair composed  
of  a `vacancy' and  an `interstitial'  is $E=0.316 
k_BT_0$. 
These defects are followed by the formation of pairs of 
extended dislocations (see Fig.~5(e) of Ref.\cite{schweigert_new}). 
As seen from Fig.~12(b) these defects do 
not change substantially either the 
bond--angular or the translational order factors.
 At $T=T_2=0.01078T_0$ the picture of defect structure
 changes crucially. Uncorrelated 
extended dislocations with non-zero Burgers vector and  
unbounded disclination pairs (see Fig.~5(e) of Ref.\cite{schweigert_new}) are formed 
which causes a substantial decrease of the translational order. These isomers 
refer to the 
next energy range up to $E \simeq 1.4k_BT_0$ (group 2).     
Then  with increasing density of 
dislocations and
the appearance  of single  disclinations the system looses 
order and  transits to an isotropic fluid 
 (group 3) in Fig.~12(b). 

Fig.~12 clearly illustrates that the energy of the different defects in both 
lattice structures is substantially different. The defects which are able to 
destroy the translational and
orientational order in the square lattice have a larger energy than 
corresponding one in the  single layer.  
Even point defects in the square lattice crystal have a  
higher energy. Thus, the square type bilayer crystal 
requires larger energies in order to create defects which are 
responsible for the loss of the bond-orientational and the translational order.

\section {Conclusion}
Monte-Carlo simulations of the melting transition of  
single and  
bilayer crystals were performed. The 
solid--liquid phase diagram  for the five lattice structures 
which are stable in the bilayer system was constructed. We found  an 
unexpected larger melting temperature for the square 
lattice. A comparison of the topology of the defects 
occurring in a hexagonal single layer crystal with these  in the square 
bilayer crystal clarifies the enhanced stability of the square 
lattice. In the case of  the square lattice, all defects have a 
larger energy and consequently larger thermal energy is 
required to create them. 

An analysis of the bond-angular and translational order factors 
of a 2D crystal for the Coulomb, screened Coulomb, Lennard-Jones 
and $1/r^{12}$ repulsive 
inter--particle interaction potentials allowed us to propose 
a new universal criterion for the melting 
temperature: {\it melting occurs when the bond-angular correlation factor 
attains the value $G_{\theta} = 0.45$}.
Using this criterion we showed that the melting temperature, 
$T_{mel}$, can be obtained with sufficient accuracy   
within the harmonic approximation.  

It is shown that the potential energy of the system 
 changes substantially  within a
 very narrow range of temperatures around the melting temperature.
 We found the equilibrium state of the system nearby the melting point where
the system alternates between the crystalline and 
the liquid state.
 The system is able to switch
 from the liquid to the solid state
 and back during the simulation time. 
Such a behaviour indicates that
 we reached the equilibrium state
 of the melting transition and 
hysteresis is absent for the melting of the
 Coulomb crystal at least for our  finite size
 (448 particles) system with periodic boundary conditions.

\section {Acknowledgments }

This work is supported by the 
Flemish  Science Foundation (FWO-Vl) and the Russian 
Foundation   for  Fundamental  Investigation 
96--023--9134a. One of us (FMP) is a Researcher 
Director with FW0-Vl.
We acknowledge discussions with G. Goldoni in the initial stage of this work.

\begin {center} { FIGURES } \end {center}

FIG. 1. The relative mean square displacement of the particles (a), the 
height of the first peak of  the pair correlation function (b), 
the potential energy (c) and the bond-angular (solid circles) and the 
translational (open circles) order factors (d) as function of 
temperature for the interlayer distance $\nu =0$.

FIG. 2. The same as Fig. 1, but now for the interlayer 
distance $\nu =0.4$, where the system forms a square lattice.

FIG. 3. The same as Fig. 1, but now
for the interlayer distance $\nu =0.8$, where the system is in a staggered 
triangle lattice. 

FIG. 4. The pair correlation function of the square
 lattice bilayer crystal ($\nu =0.4$) at $T=0.0005T_0$ (solid curve),
 $T=0.0095T_0$ (dotted curve) just before melting and $T=0.011T_0$ (dashed 
curve) after melting where
the system is in the liquid state.

FIG. 5. The phase diagram of the bilayer Coulomb crystal 
for different screening lengths $\lambda =0$ (open squares), 
$\lambda =1$ (solid circles), and $\lambda =3$ (open triangles).
The curves are guides to the eye. The crystal structures are shown in the 
inserts where open (solid) symbols are for the particles in the top (bottom) 
layer.

FIG. 6. The structural phase diagram of the bilayer 
screened Coulomb crystal at $T=0$ as function of the 
screening length $\lambda $.

FIG. 7. The potential energy $(E(T)-E(T=0))/k_BT_0$ (a) and the bond--angular order 
factor (b) as 
function of temperature for the interlayer distances: $\nu =0$ (solid squares), 
$\nu =0.4$ (open triangles), for the screened Coulomb inter-particle 
potential with screening length $\lambda =1$.

FIG. 8. The translational (a) and the bond-angular order 
factors (b) as function of temperature for the interlayer 
distances $\nu =0$ ($\lambda =1$--solid squares; $\lambda =3$-open circles), and 
$\nu =0.4$ 
($\lambda =1$--solid triangles; $\lambda =3$--open triangles), 
for the Lennard-Jones potential (solid rhombics) and for 
the repulsive potential $1/r^{12}$ (open rhombics).

FIG. 9. The melting temperature in the harmonic 
approximation (solid curves) and from MC simulations 
(symbols): (a) for the square bilayer crystal as function of 
the interlayer distance for screening lengths $\lambda =0$ 
(circles), $\lambda =1$ (squares), $\lambda =3$ 
(triangles), and (b) for the hexagonal single layer crystal as 
function of the screening length.

FIG. 10. Chain of dislocations (grain boundary): (a) in 
the vertical direction, and (b) in the horizontal direction in a single layer 
hexagonal 
lattice. 
 
FIG. 11. The bond-angular (squares) and the translational 
(open circles) order factors (a) and $(E(T)-E(T=0))/ k_BT_0$ (b) as function of  
`MC time' steps.

FIG. 12. The bond-angular (squares) and the translational 
order (circles) factors  of the different defects in (a) a single 
layer crystal $\nu =0$, and (b) in the square lattice bilayer 
system for $\nu =0.4$. 

FIG. 13. Defects in a single layer crystal: (a) `centred 
vacancy' and (b) `centred interstitial'.

FIG. 14. The point defects: (a) `twisted bond', (b) `twisted 
triangular', (c) 'twisted square', (d) `vacancy', and (e) 
`interstitial' in the square lattice bilayer crystal. 

\newpage
\end {document}
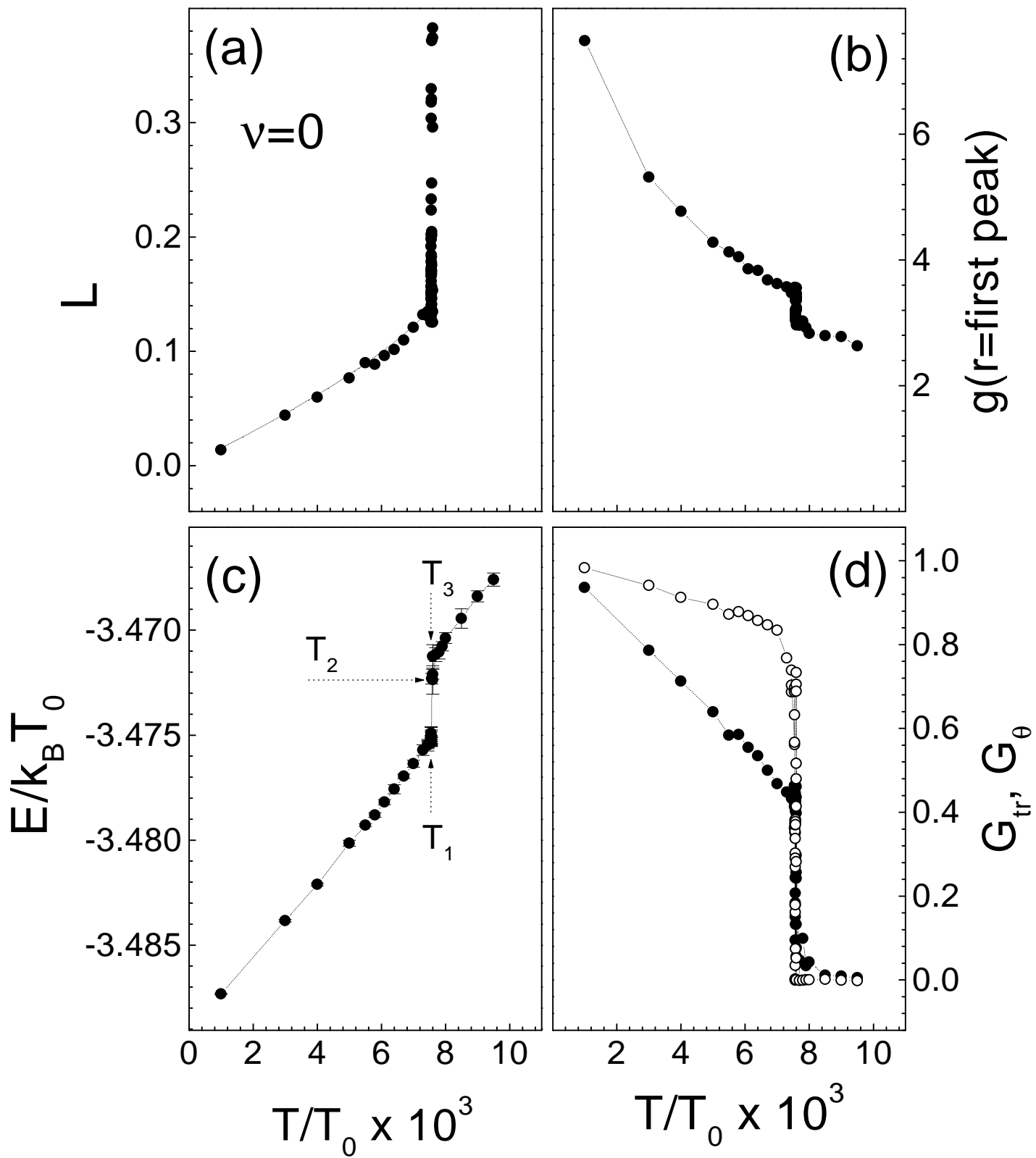
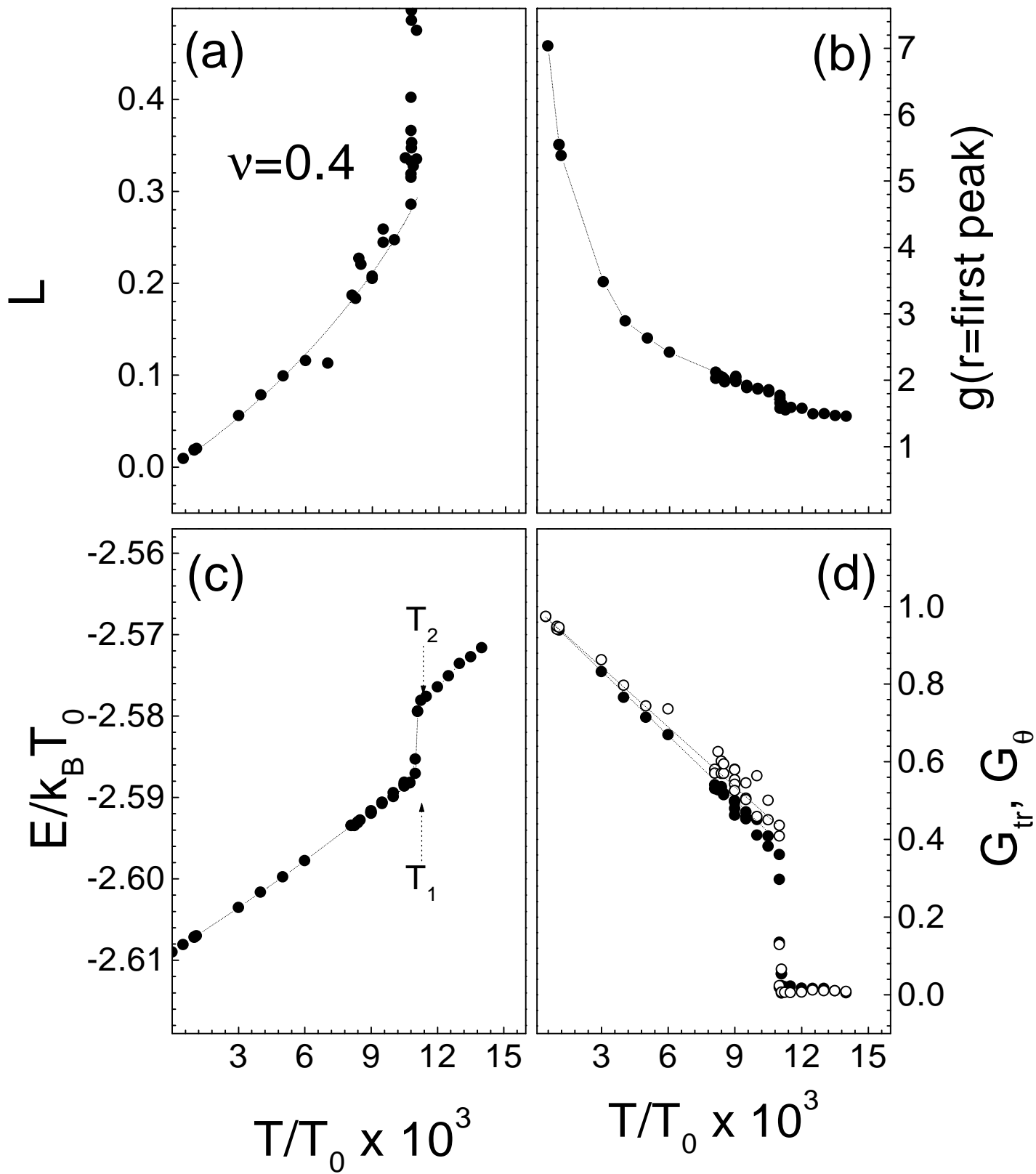
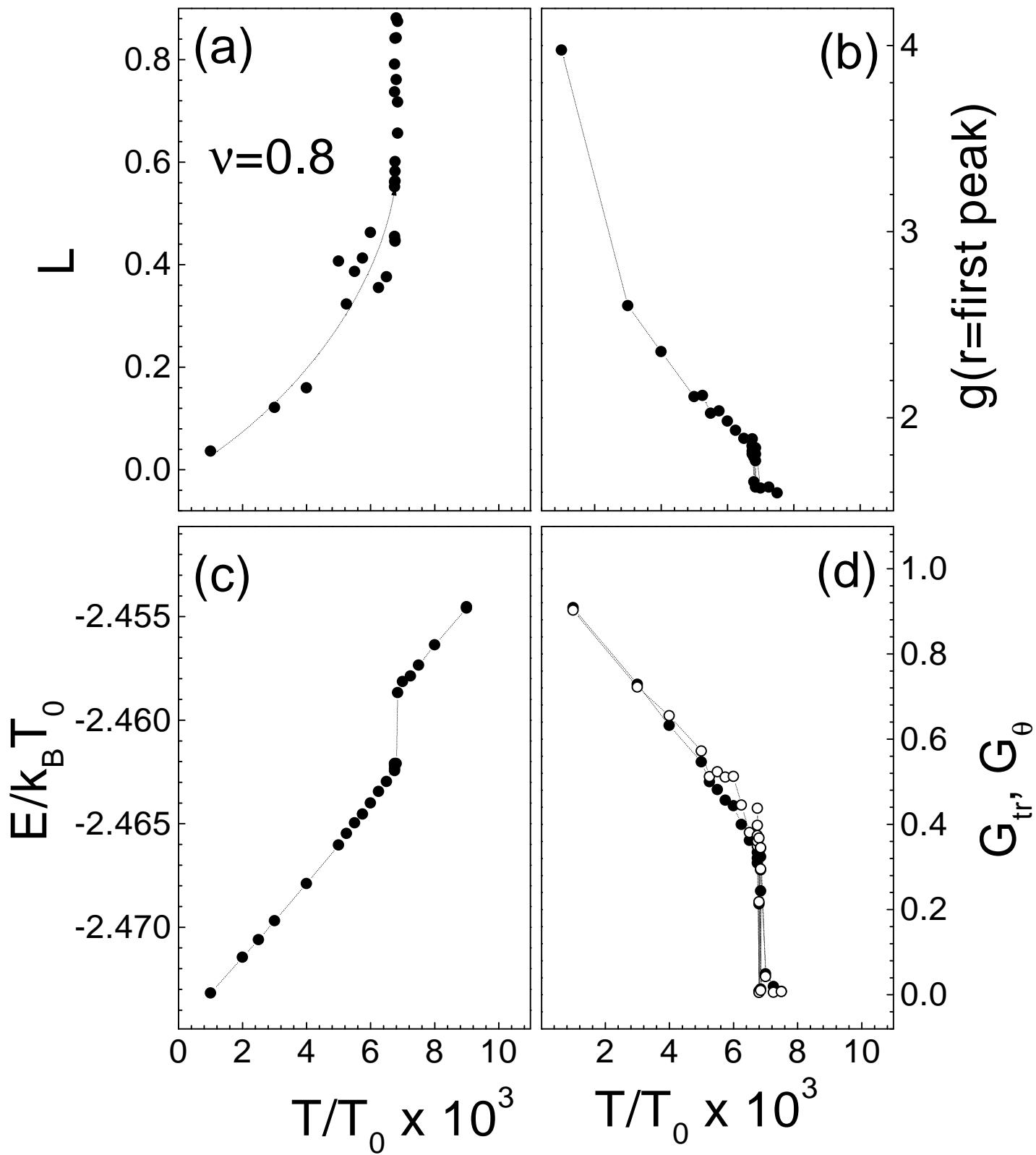
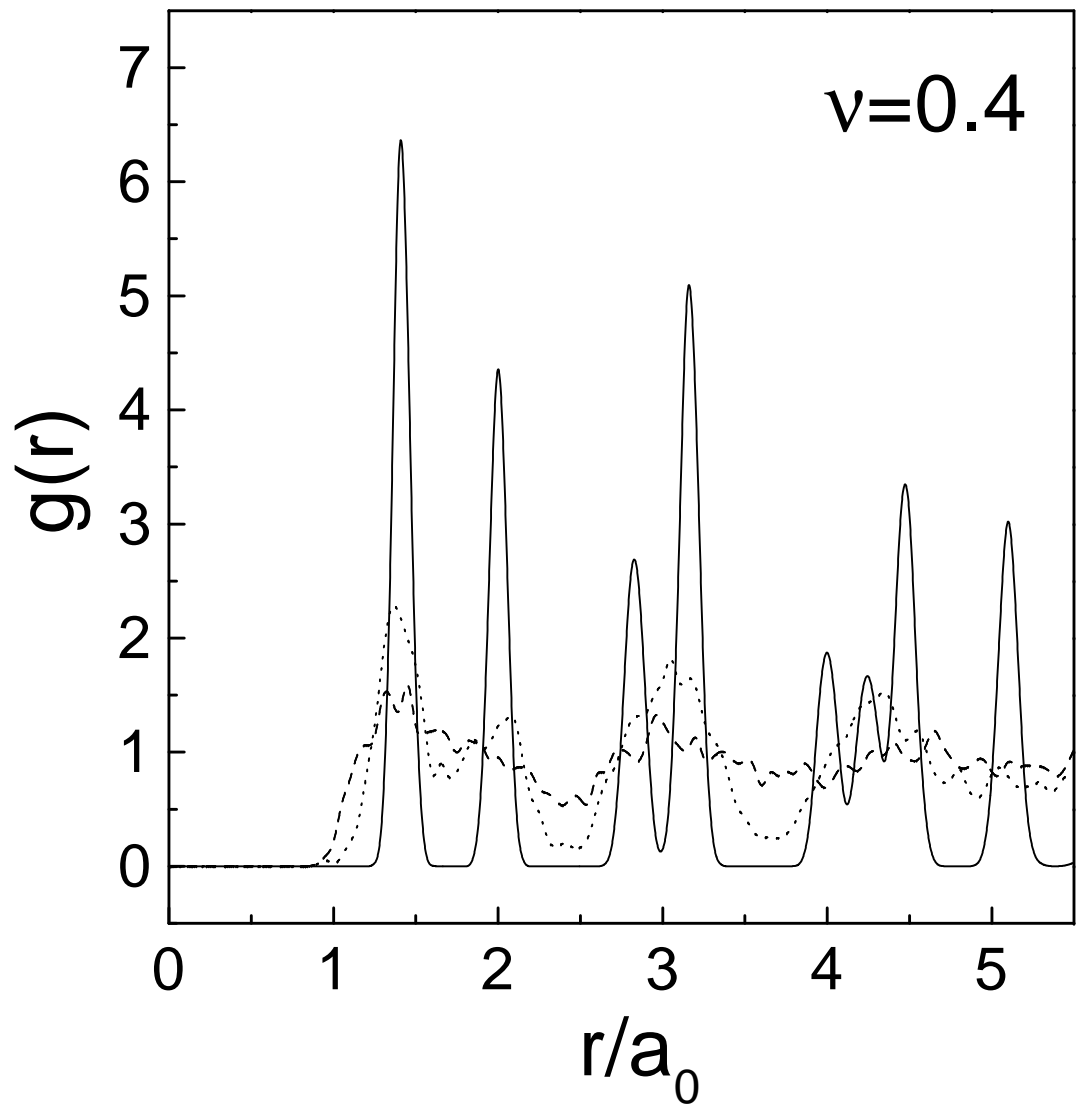
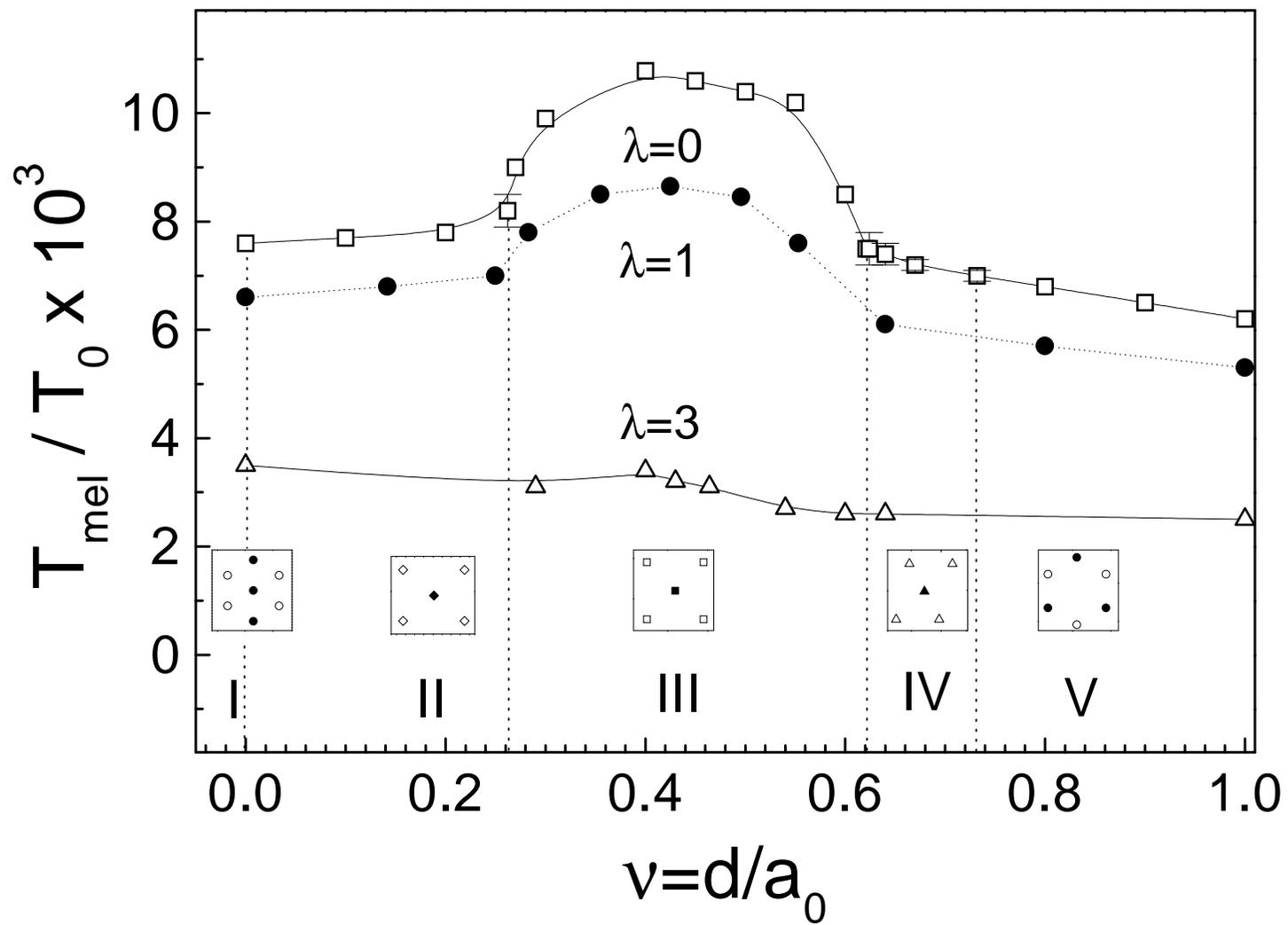
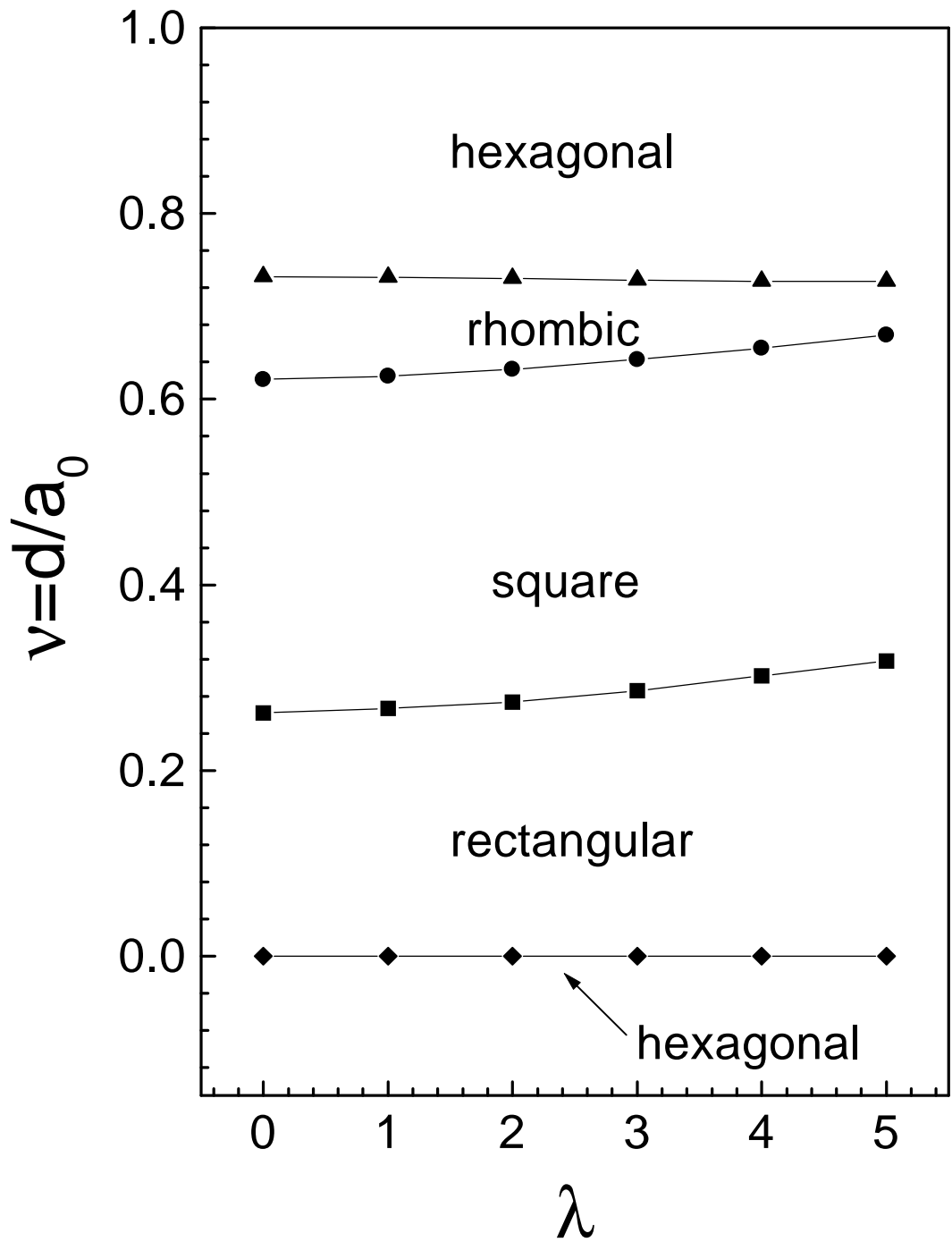
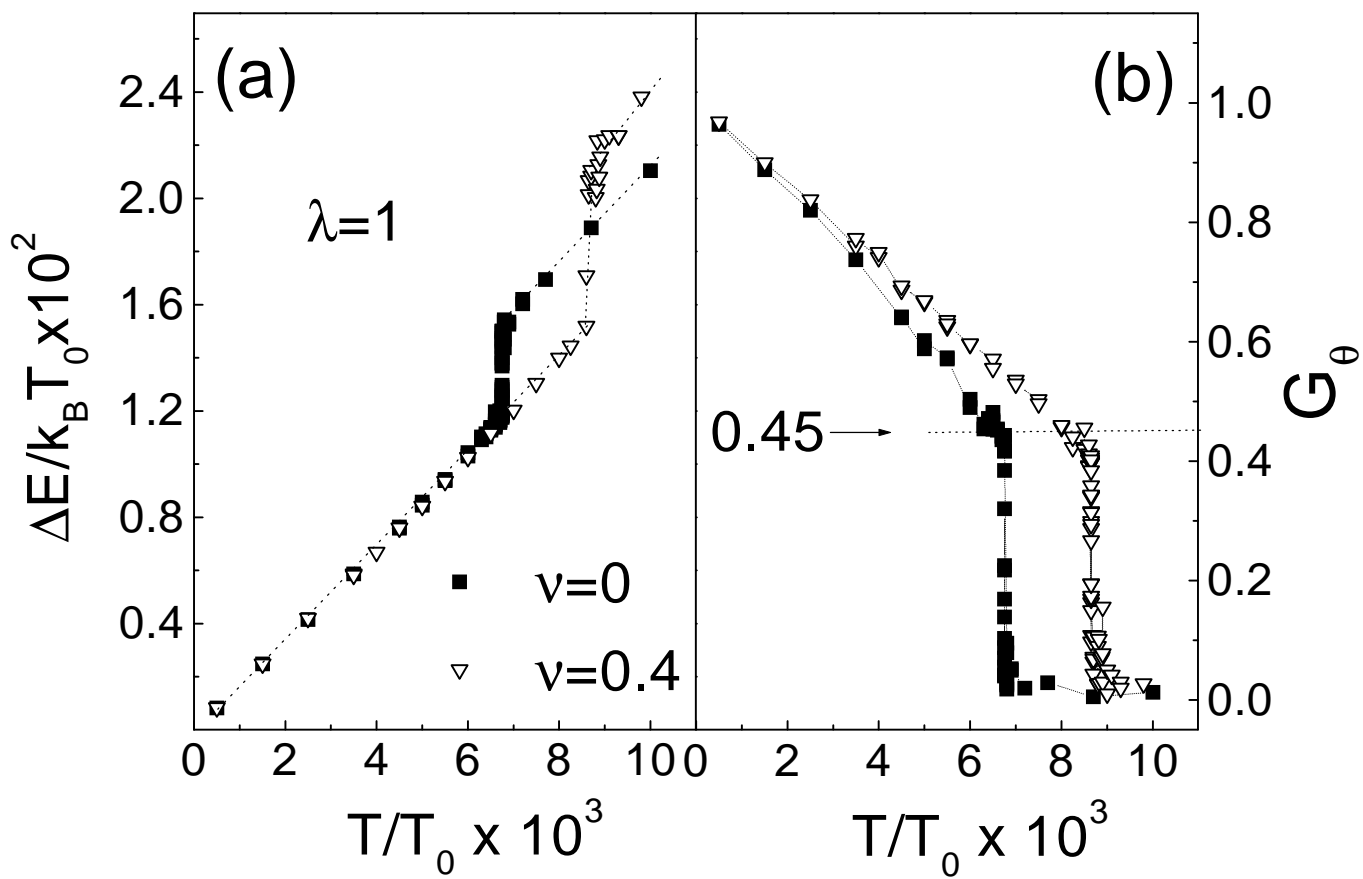
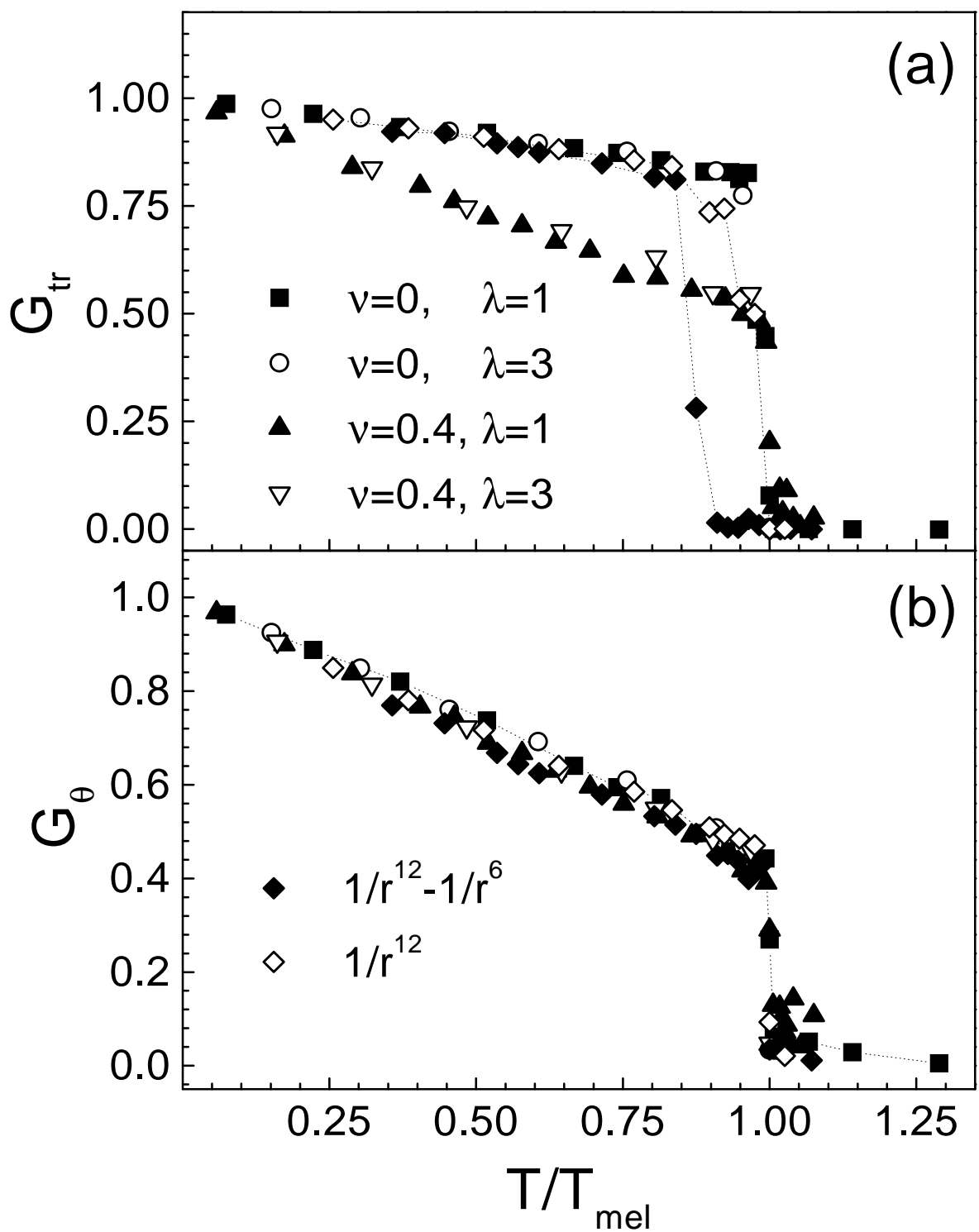
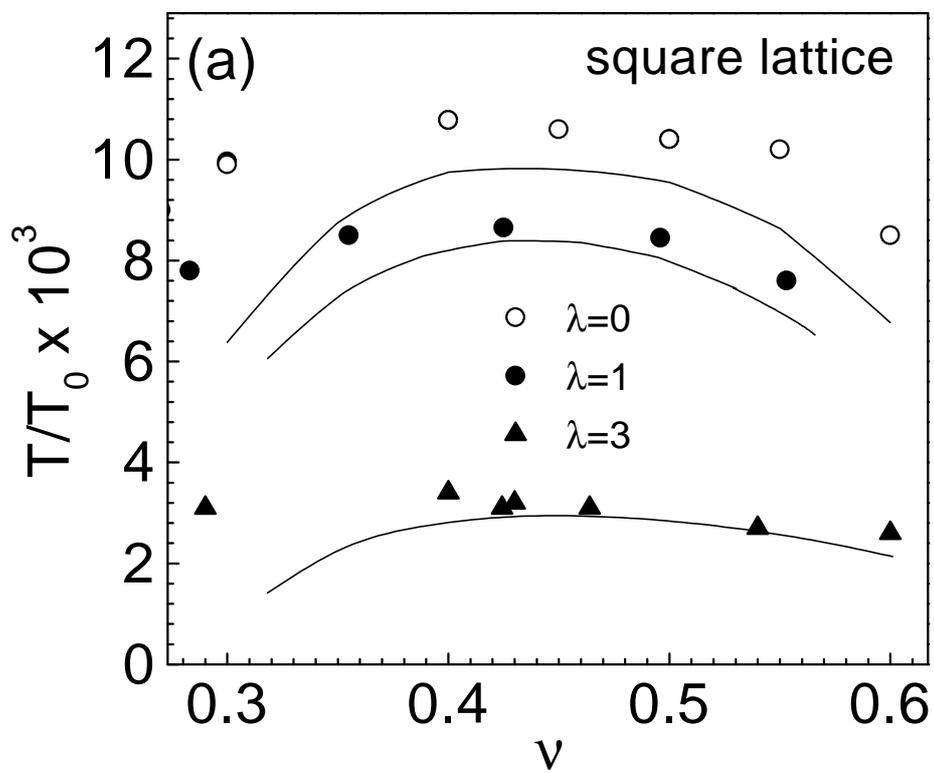
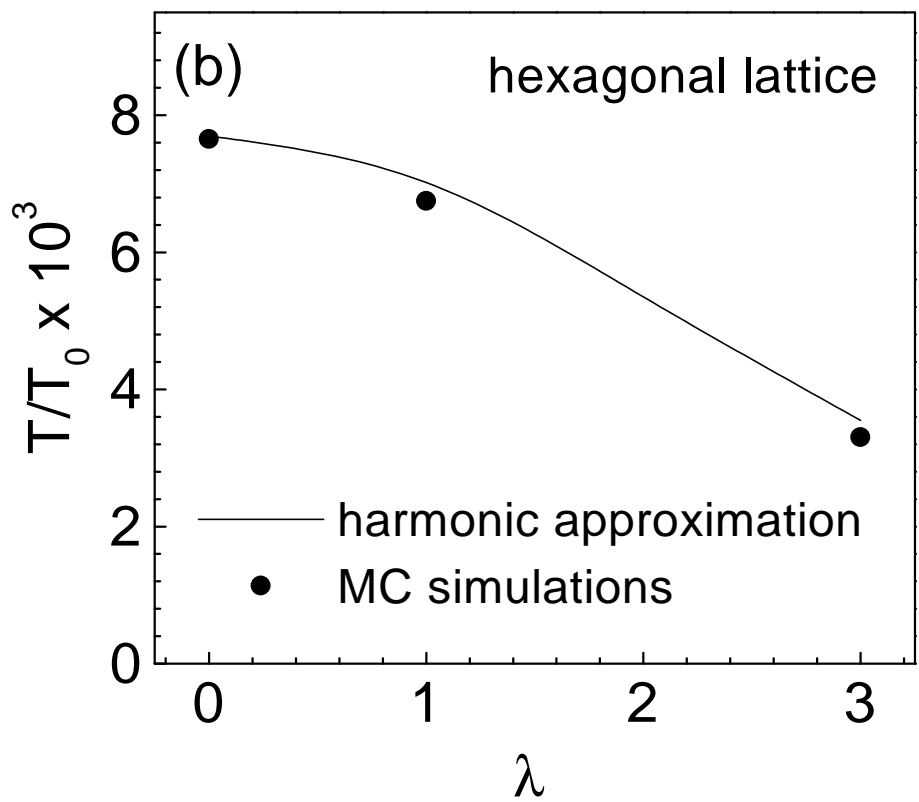
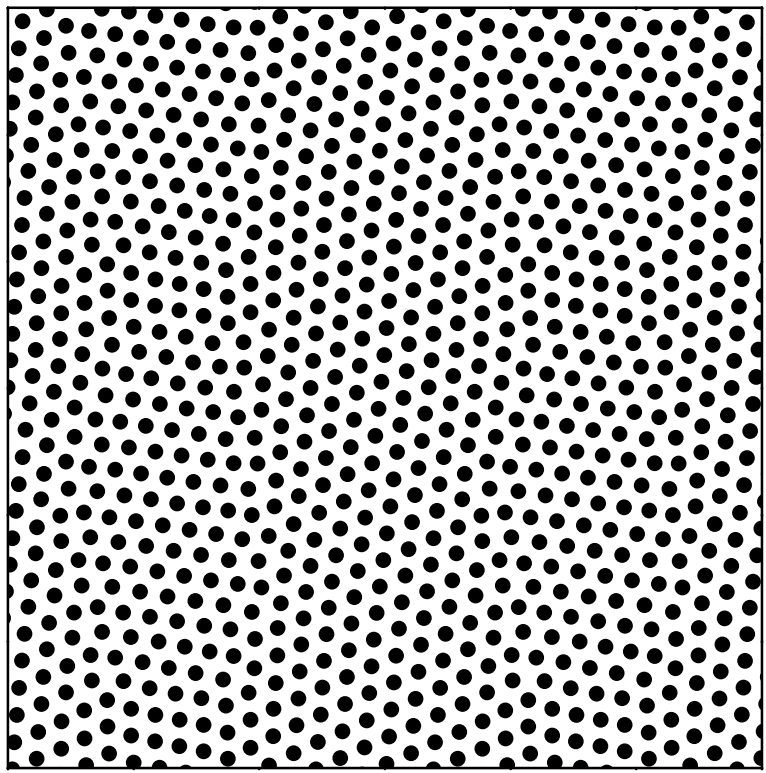
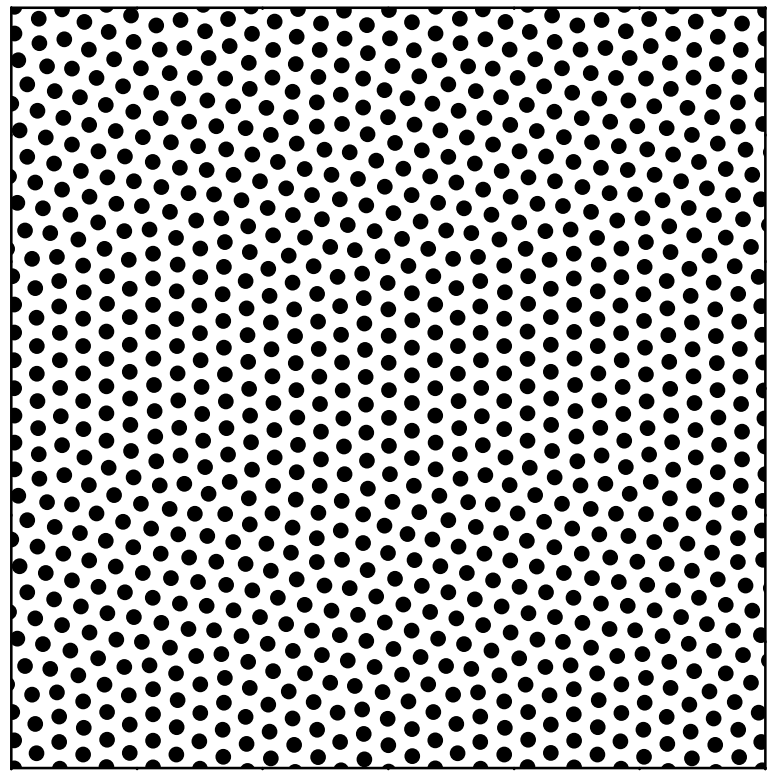
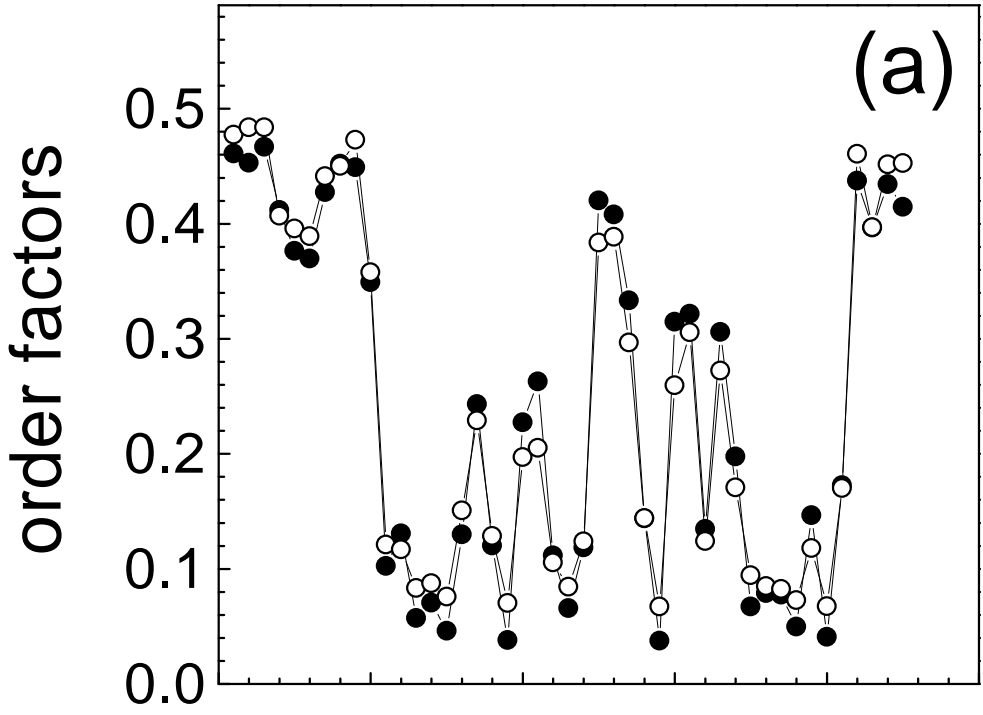
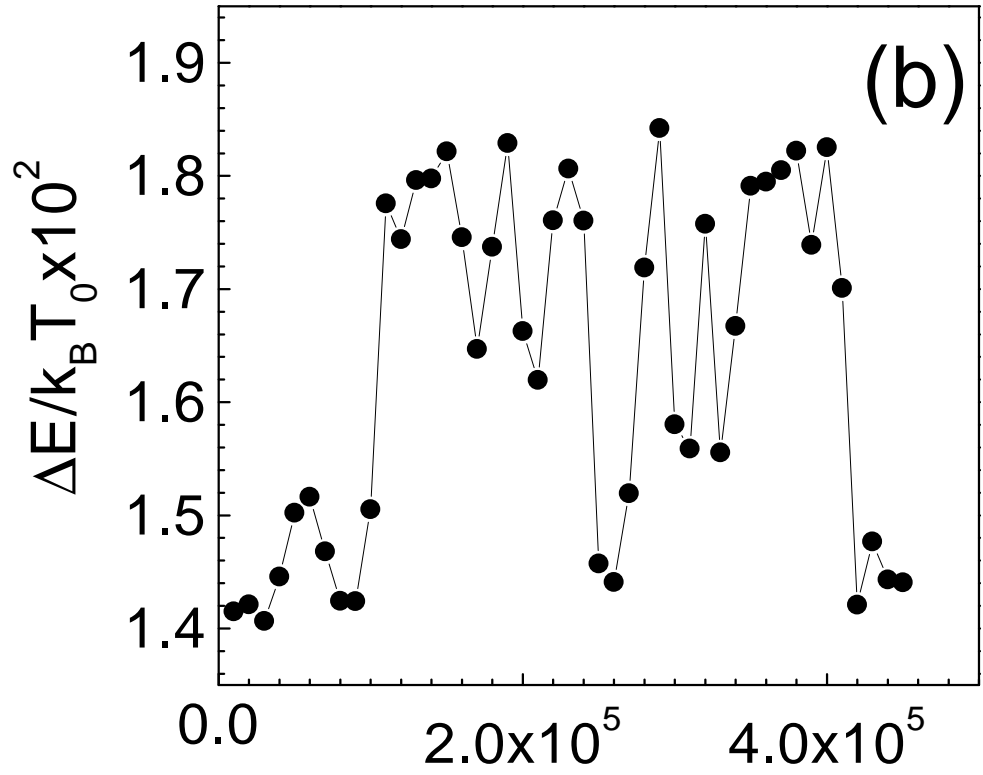
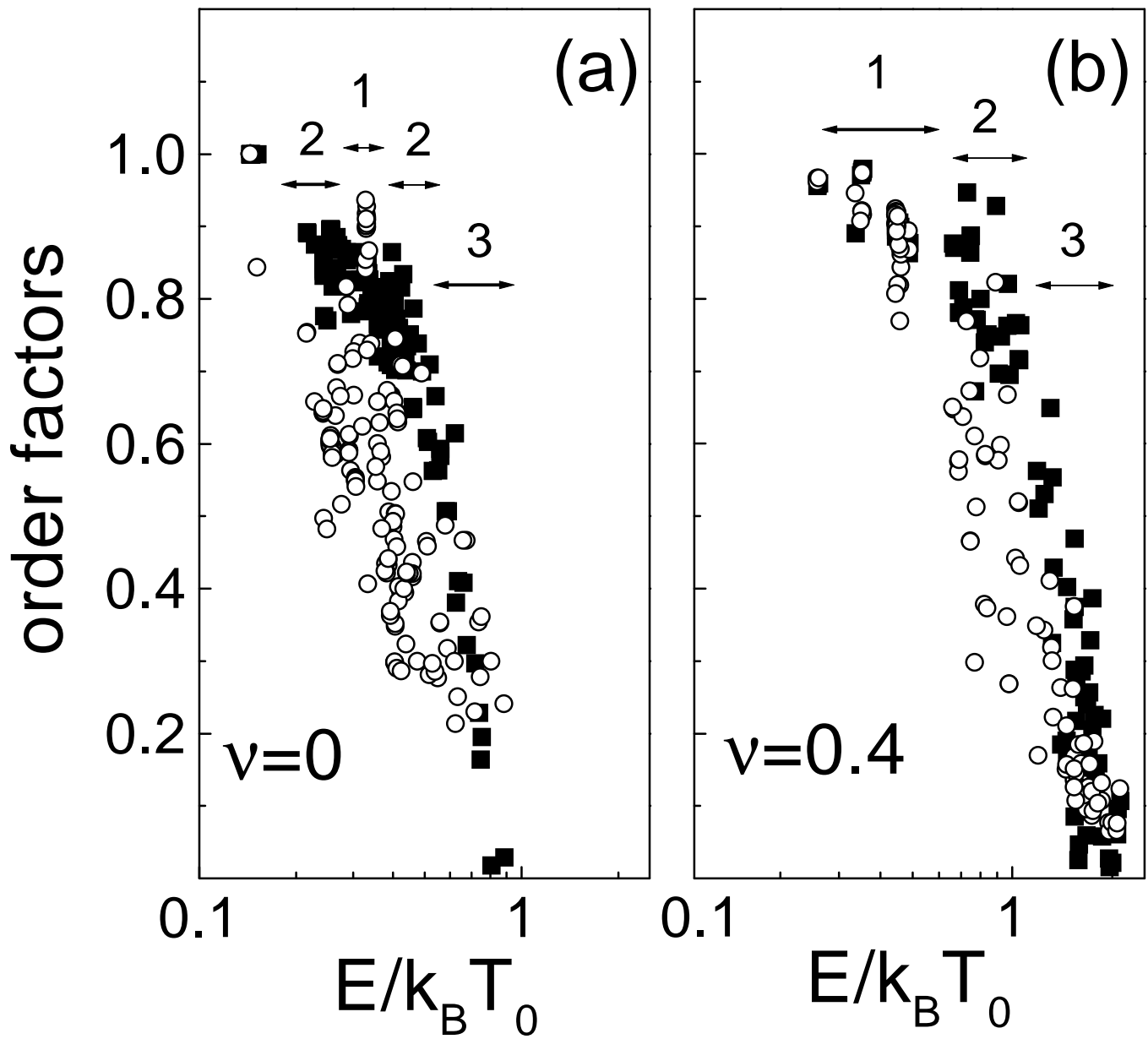
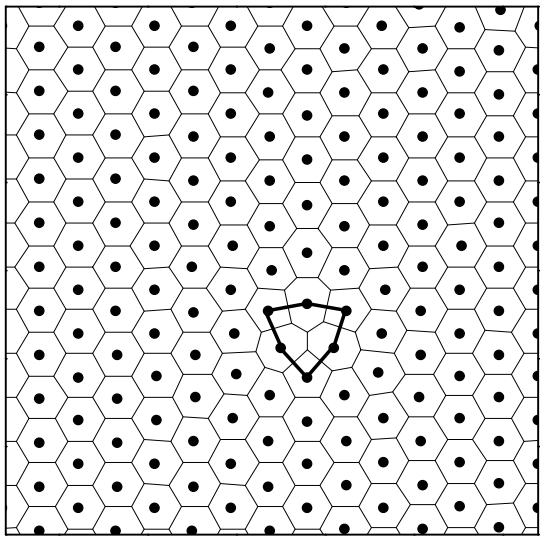
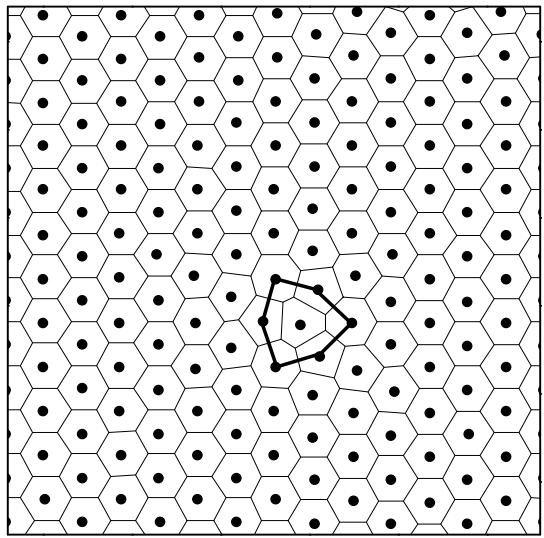
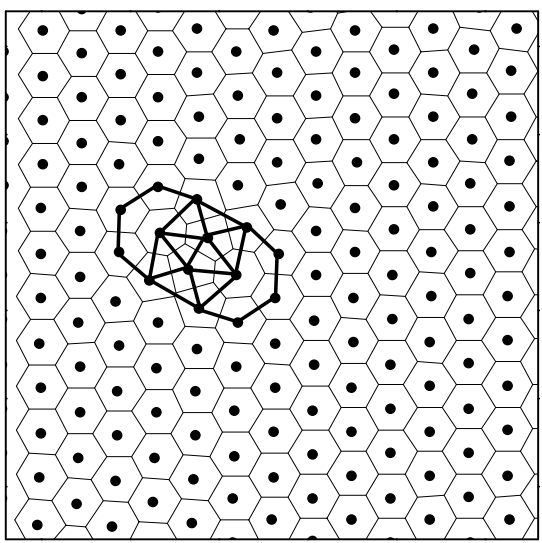
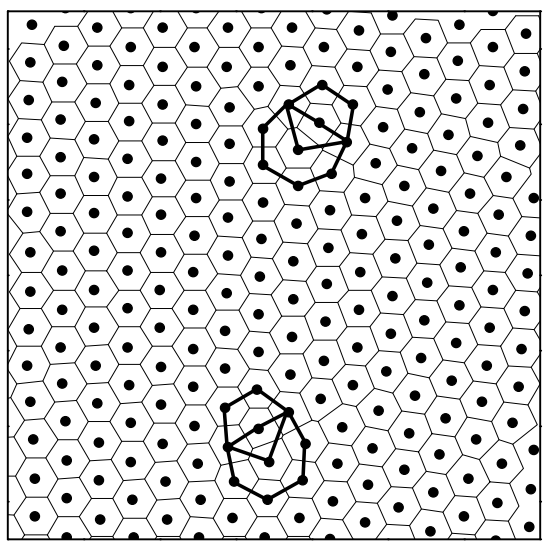
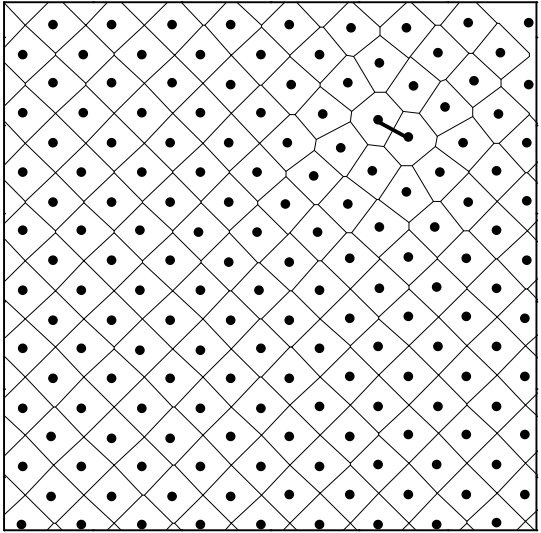
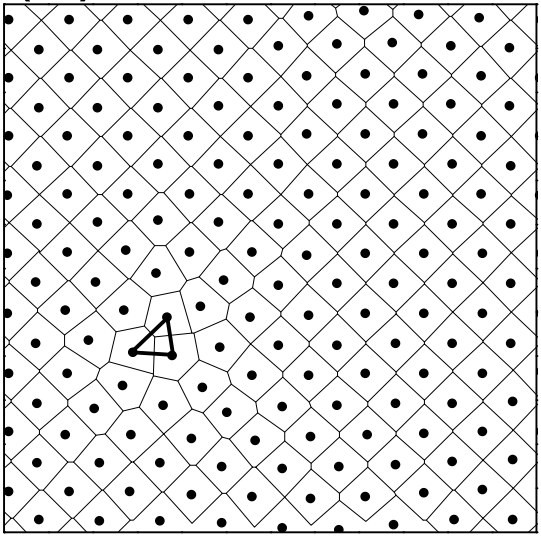
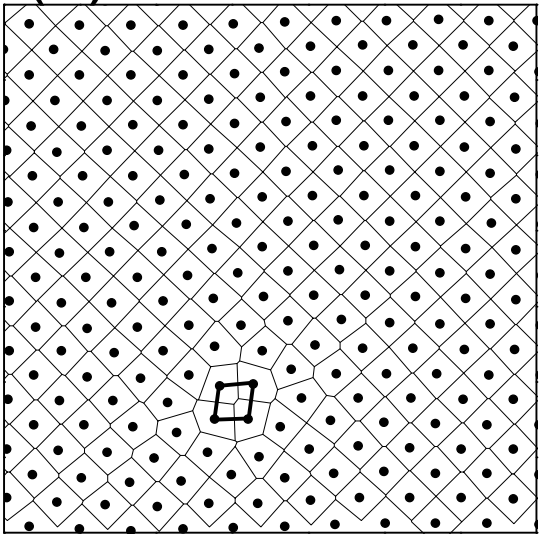
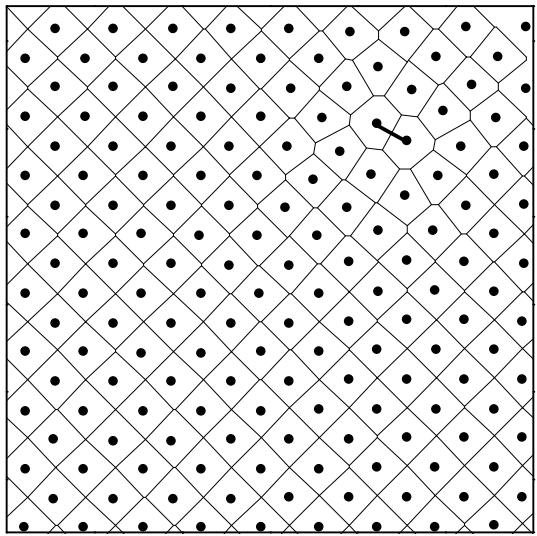
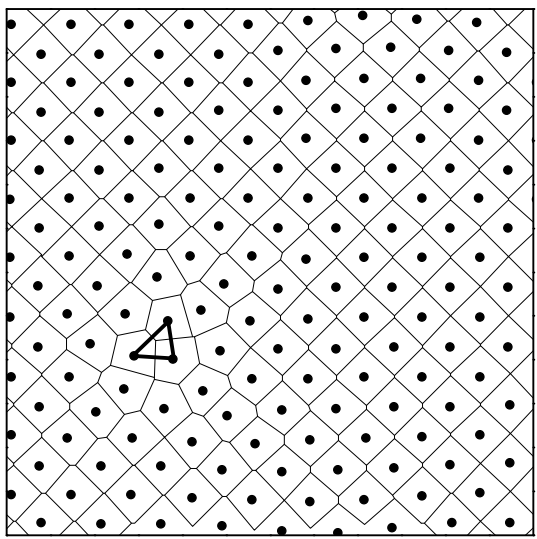
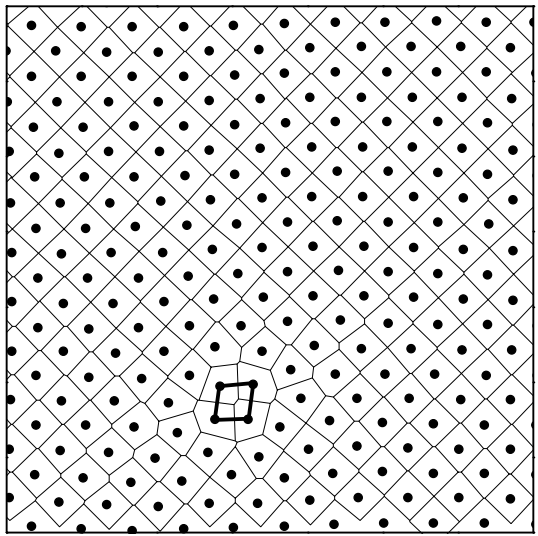
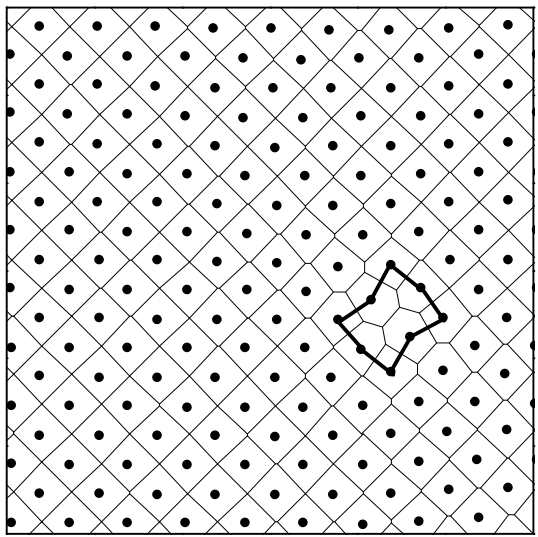
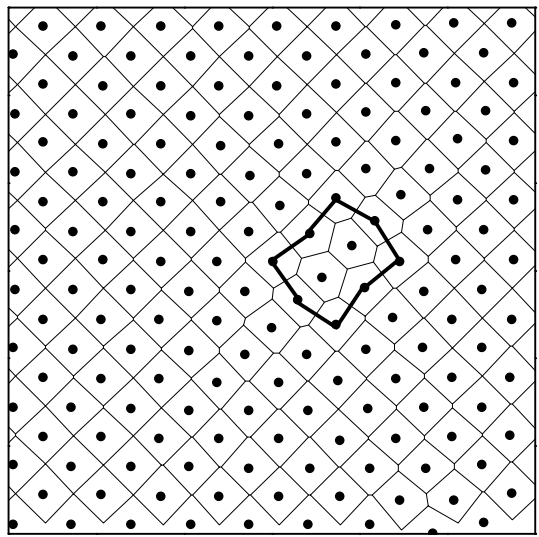
